\documentclass[conference]{IEEEtran}
\IEEEoverridecommandlockouts
\usepackage{cite}
\usepackage{amsmath,amssymb,amsfonts}
\usepackage{algorithmic}
\usepackage{graphicx}
\usepackage{xcolor}
\usepackage{listings}
\usepackage{textcomp}
\usepackage{xcolor}
\usepackage{subcaption}
\usepackage{url}

\definecolor{MyRed}{rgb}{1,0,0}

\newcommand{\ellide}[1]{}

\def\BibTeX{{\rm B\kern-.05em{\sc i\kern-.025em b}\kern-.08em
    T\kern-.1667em\lower.7ex\hbox{E}\kern-.125emX}}
\begin{document}

\title{CkIO: Parallel File Input for Over-Decomposed Task-Based Systems
}

\author{
    \IEEEauthorblockN{Mathew Jacob, Maya Taylor, Laxmikant Kale}
    \IEEEauthorblockA{University of Illinois at Urbana-Champaign
    \\\{mkjacob3, mayat4, kale\}@illinois.edu}
}

\maketitle

\begin{abstract}
Parallel input performance issues are often neglected in large scale parallel applications in Computational Science and Engineering. Traditionally, there has been less focus on input performance because either input sizes are small (as in biomolecular simulations) or the time doing input is insignificant compared with the simulation with many timesteps. But newer applications, such as graph algorithms add a premium to file input performance. 
Additionally, over-decomposed systems, such as Charm++/AMPI, present new challenges in this context in comparison to MPI applications. In the over-decomposition model, naive parallel I/O in which every task makes its own I/O request is impractical. Furthermore, load balancing supported by models such as Charm++/AMPI precludes assumption of data contiguity on individual nodes. We develop a new I/O abstraction to address these issues by separating the decomposition of consumers of input data from that of file-reader tasks that interact with the file system. This enables applications to scale the number of consumers of data without impacting I/O behavior or performance. These ideas are implemented in a new input library, CkIO, that is built on Charm++, which is a well-known task-based and overdecomposed-partitions system. CkIO is configurable via multiple parameters (such as the number of file readers and/or their placement) that can be tuned depending on characteristics of the application, such as file size and number of application objects. Additionally, CkIO input allows for capabilities such as effective overlap of input and application-level computation, as well as load balancing and migration. We describe the relevant challenges in understanding file system behavior and architecture, the design alternatives being explored, and preliminary performance data.

\end{abstract}


\section{Introduction} \label{Introduction Section}
I/O-bound tasks are important throughout high performance computing, from writing the current state of the program to disk during a checkpoint or reading in particle data to start a large simulation. Party due to the disk and file systems being orders of magnitude slower than the rest of the system, these I/O tasks often become a bottleneck. Thus, many researchers have focused on optimizing I/O to improve application performance. This existing research, however, tends to focus on optimizing the output performance rather than the input, and traditionally input performance hasn't been a major concern \cite{Charm_output} \cite{MPIIO}. For modern workloads, however, input performance is becoming increasingly important. Programs, such as
graph applications, typically run for relatively short time, and may be time-critical, increasing the importance of fast reading time. Even for large-scale iterative scientific applications, such as N-body astronomy simulations, which can run for hours, completing many iterations and time steps, performance tuning at scale requires short runs with only a few timesteps; in these situations, poor input performance becomes a pressing concern. 

\begin{figure}

  \centering
  
 \includegraphics[width=.8\linewidth]{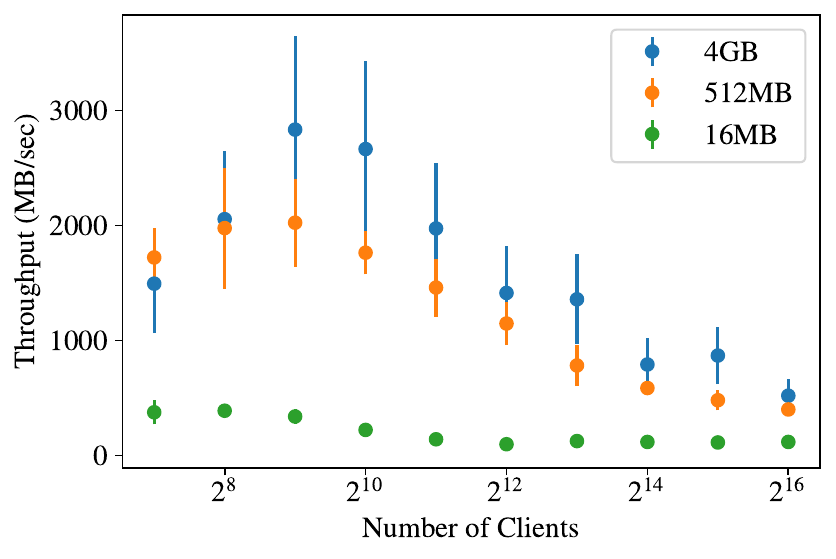}
  \caption{Naive overdecomposed input in Charm++. Results were produced on Bridges2 using 16 nodes and 512 PEs, where each data point denotes an average over 10 runs.}
  \label{fig:naive-bridges-tp}

\end{figure}

Asynchronous Many-Task (AMT) systems, such as Charm++ \cite{Charm++}, HPX \cite{HPX}, as well as the task-based implementation of Legion \cite{legionMain}, decouple the input decomposition from the number of processing elements (PEs). This decoupling allows for \textit{overdecomposition}, meaning that multiple parallel tasks can be assigned to a single PE, and the runtime is responsible for making these assignments. Overdecomposition, combined with adaptive dynamic scheduling of tasks depending on availability of data, can improve performance by increasing the overlap of computation and communication in a program. However, in the context of I/O, this overdecomposition can degrade performance substantially if it is not tuned to the file system. This phenomenon is illustrated via the following experiment. 
We use the term \textit{clients} to refer to the overdecomposed tasks that perform I/O. In this experiment each client {\em directly} (i.e. by making in individual file system call) reads a disjoint section of a single input file; we use 3 different file sizes and vary the number of clients while keeping the number of nodes and PEs constant as we measure the throughput (i.e. the file input rate, higher the better).
For each [filesize,number-of-clients] configuration, we repeat the experiment multiple times because of observed variability, and indicate the extent of variability by the vertical bars.
The results plotted in Figure \ref{fig:naive-bridges-tp} show that the input performance is significantly dependent on the number of clients: if the number of clients is too small, performance suffers because the program misses opportunities for parallelism in disk accesses, and if the number of clients is too large, the file system is congested by many small read requests. 

Yet, from the application developer's perspective, the number of clients should not be constrained by the optimal I/O decomposition. The user will choose the number of clients per PE (overdecomposition factor) based on what is optimal for the computational phases of a given application; E.g. in ChaNGa, a computational astronomy application, it is common to see 16 objects per core \cite{CHaNGa}. So, an important issue for I/O performance in overdecomposed programming models is how to allow the user complete freedom over their application decomposition and the number of clients, while supporting consistent, optimal I/O performance. This is the first issue this paper addresses.

Additionally, to retain the performance benefits of the data-driven task-based execution model, it is important to ensure that I/O operations and unrelated tasks can be scheduled together. 
At the same time, tasks that do depend on specific input data should be enabled (i.e. marked ready for scheduling) as soon as the data is available. 
This requires supporting non-blocking split-phase interfaces for I/O operations, 
where triggering of each input call is separated from its completions and the continuation of dependent tasks. 

Futhermore, in overdecomposed systems such as Charm++, tasks are able to migrate across processors/nodes during execution. A task may open a file on one node, read some initial data and then get migrated to another node, where it continues to read from the same open file handle, further complicating I/O in this context.

\begin{figure}

  \centering
  
 \includegraphics[width=1.1\linewidth]{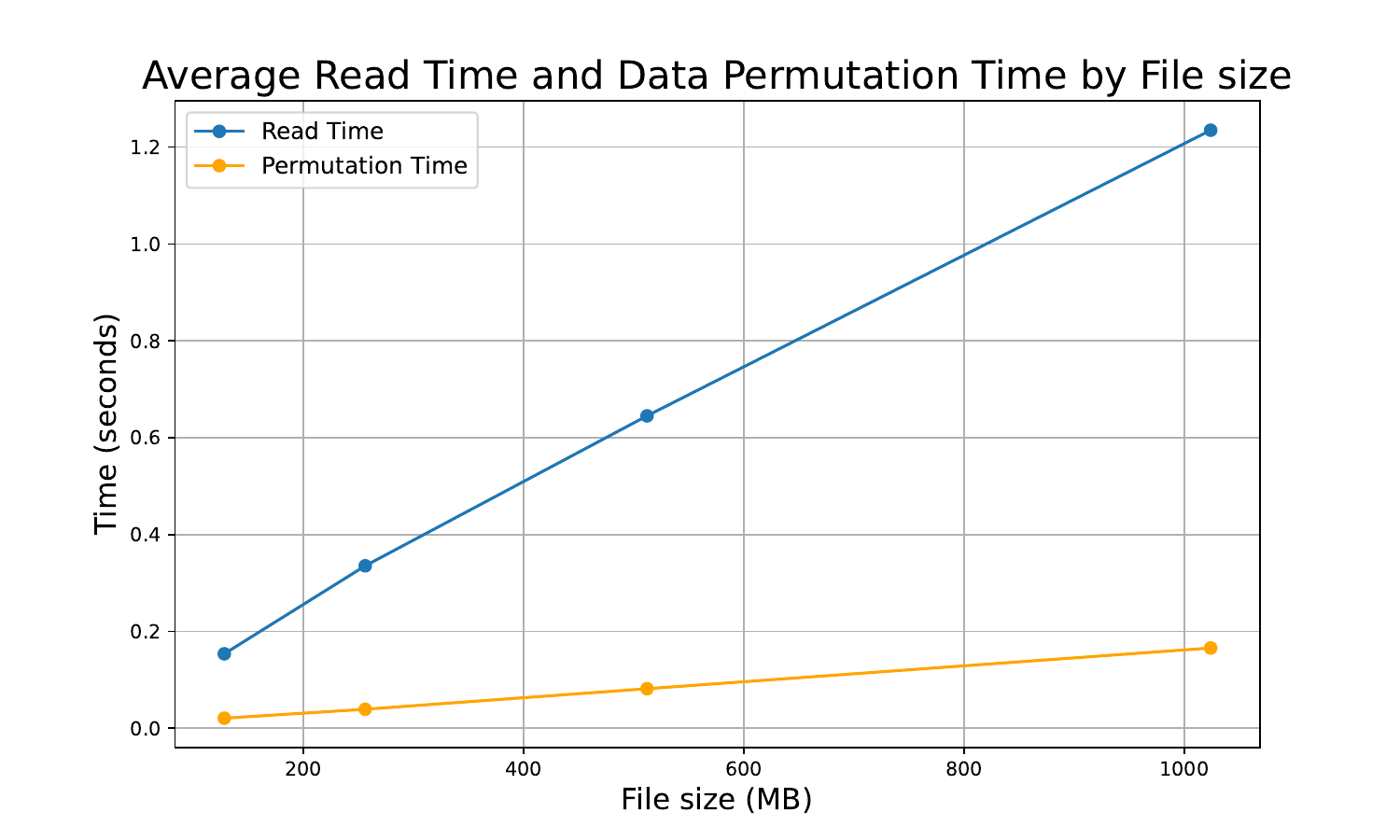}
  \caption{This graph compares the time to read data from the filesystem and sending that data across nodes. The x-axis denotes the size of the file being read and sent over the network, while the y-axis is the time.}
  \label{fig:permutation_io}

\end{figure}

To approach the issues described, we begin by noting that on modern parallel clusters, reading data from disk is an order of magnitude slower than moving that same data across nodes within the parallel computer. This is substantiated by experimental data shown in Figure \ref{fig:permutation_io}, which shows the performance contrast between reading different file sizes from the file system and transmitting that same data over the network from one task to another. This experiment used 2 Regular Memory (RM) nodes on Bridges2\cite{Bridges2}, each with one task per node. To measure the I/O latency, one task read in the entire file. To measure the network bandwidth, one task sent the data from the file over the network to another task on a different node. This experiment illustrates that transmitting the same data over the network can be over 6x faster than retrieving that data from the file system in task-based systems on parallel-computing clusters. This allows us to consider solutions where data requested by a client on one processor can be read from the file system by another processor on another node. 

We design CkIO, an input scheme for over-decomposed, task-based systems, based on two-phase I/O \cite{twophase}.
CkIO introduces a configurable intermediary layer of objects between the file system and application tasks. The intermediary tasks themselves use asynchronous input via helper pthreads to carry out reads for large aggregated chunks of data, and
then fulfill requests from application tasks by sending the corresponding data over the interconnects, which are much faster than the I/O subsystems on most parallel computing clusters. Additionally, our API design as well as its implementation allows for effective overlap of the I/O and computational work of applications. We implement this scheme in Charm++, an asynchronous task-based parallel programming framework \cite{Charm++}. 

This paper makes the following contributions:
\begin{itemize}
    \item A new approach for optimizing file input performance in over-decomposed task-based systems. One key idea is designate a subset of tasks to greedily read in a \textit{read session}, which is a user-specified section of the file that the user will read in a given phase, asynchronously. Combined with an "asynchronous callback architecture, described in the paper, this effectively overlaps I/O with non-dependent computation, while still achieving the best raw input throughput.
    \item An open source implementation of the approach in CkIO, a parallel input library for overdecomposed systems, as a part of Charm++.
    \item We show that CkIO achieves performance competitive with MPI I/O even when no overdecomposition or task-based scheduling is involved and achieves superior performance in their presence, via input aggregation and overlap of both application-level computation and I/O. 
    We demonstrate this by integrating CkIO into ChaNGa, an N-body gravity simulator, where CkIO provides over a 2$\times$ speed-up, compared with the hand-optimized I/O code that ChaNGa previously used. 
    \item We demonstrate that our scheme supports migratability in I/O, i.e. allows persistent tasks (chare objects) to migrate across nodes while holding an active file handle and continuing read operations from multiple places. 
\end{itemize}

\section{Background} \label{Background Section}

\subsection{Overdecomposition}
\textit{Overdecomposition} breaks up a program into many smaller domains, often in the form of units of data and work or computation. This subdivision is independent of the number of processors being used, and separates the work done by the application from the actual hardware of the resources used to run the application. This gives the runtime the ability to schedule computational work as it sees fit, which often leads to high performance via automatic optimizations, that would otherwise be challenging for a programmer to achieve. This programming paradigm also allows for the runtime to overlap computation and communication, which further improves the performance of overdecomposed applications.

A related category is task-based systems, or systems that support message driven execution. They exhibit similar features and require similar support for parallel I/O. Notable such systems including Legion \cite{legionMain}, HPX\cite{HPX}, and StarPU\cite{StarPU}.

\subsection{Charm++}
Charm++ is a task-based, parallel programming framework that enables overdecomposition and other related optimizations, such as dynamic load balancing \cite{Charm++}. A Charm++ program is decomposed into work units called \textit{chares}.

\subsubsection{Chares, Chare Arrays, and Groups}
Chares are the fundamental building block of a Charm++ program. These chares are  C++ objects that the runtime assigns to processors in a system. A chare encapsultes its own data, and is only allowed to access and update its own data, although it can access global readonly data and can hold handles (proxies) to other chares. These handles can be used by chares to send messages to other chares in the form of asynchronous method invocations. 
It gives rise to a {\em series} of atomic (i.e non-preemptible) tasks associated with its owned data based on dependencies on messages sent to it, as well as program order dependencies. (For brevity, we sometimes refer to a chare itself as a task or ``persistent task''). 
A \textit{chare array} is an indexed collection of chares that is distributed across the various processors under the control of the Charm++ runtime system. The elements of chare arrays can be migrated across processors and nodes dynamically, which supports dynamic operations such as load-balancing. A \textit{chare group} is a collection of chares such that there is exactly one chare per processor.  The chare groups are typically used for system functionalities and services such as file input. Other chares on a given processor can access a chare group member on that processor via a direct sequential object pointer, without the need to employ asynchronous method invocations.

\subsubsection{Asynchronous Message-Driven Execution}
The asynchronous method invocations meant for the chares are the {\em tasks} that are scheduled by a user-space scheduler in Charm++ runtime. Some tasks may also represent user-level (run) threads associated with a chare that are ready to execute. 
Charm++  follows an implicit and asynchronous execution model, where there is no explicit guarantee on the ordering of message execution. This simply means that the runtime is free to select any of the ready tasks for execution at each scheduling point.
As a corollary, no task is allowed to block the processor, and execution proceeds based on availability of data.
This allows the runtime to have more freedom when scheduling work and can allow for greater overlap of computation and communication, which results in greater speed ups of application, especially in an over-decomposed application.  

Our parallel input scheme caters to the file input requests coming from elements of chare arrays as clients (which are primary constructs used by applications); Further, our library also leverages chare groups and chare arrays as implementation mechanism to support our design goals.



\subsection{Related Work}

There is a significant body of research on parallel input and output, both in the context of MPI as well as some task-based systems.

In the MPI framework, the MPI-IO library supports both parallel input and output for high-performance computing (HPC) applications \cite{MPIIO}. MPI-IO provides an interface for collective, two-phase I/O and supports non-blocking variants of I/O operations; asynchronous reads, for example, are supported via the  \textit{MPI\_File\_iread\_at} API. However, given the structure of MPI programs, it can be difficult to use this interface to effectively overlap different computation and I/O in MPI. MPI also lacks support for pre-fetching during file input, and doesn't handle migration or over-decomposition.


Legion \cite{legionMain} is a parallel programming system with a task-driven runtime and support for data-driven task graphs. 
Within Legion,  I/O tasks are handled by the Iris system \cite{LegionInput}. Iris aims to hide the latency of I/O operations by creating an abstraction known as \textit{external resources}, which is meant for resources outside the life of the program, such as files. Legion also provides semantics that have different rules regarding who can modify/have access to certain regions, allowing for optimizations such as creating a local copy to avoid many disk accesses. Iris supports concurrent computation and I/O via a deferred execution model in which the Legion runtime aims to overlap communication and computation when possible. However, to the best of our knowledge, Iris does not support input optimization via aggregation and collective I/O calls, which can result in better raw I/O performance on file systems of parallel computing clusters. Another important distinction between Legion/Iris and CkIO is that we support input in presence of object (task) migration across nodes, which is essential for systems such as Charm++ which rely on migration for runtime adaptivity.


Previous work includes support for parallel output in Charm++ \cite{Charm_output} File output is typically simpler than input (although more important) because there are typically no computational tasks that are dependent on completion of file output. When there are buffer dependencies (i.e. the source buffer for file output needs to be reused after output is complete), they can be settled by lax time-step barriers.

Another category of research focuses on separate file I/O, wherein every process or I/O participant operates on a unique file. This approach is commonly seen in MPI programs and is supported by MPI-IO. Our focus in this paper is on situations when all relevant data is in a single large file, to be collectively read by a collection of tasks.

Hierarchical data format (HDF) and other data compression variants are commonly used in HPC applications and supported by many I/O frameworks \cite{hdf5}. These are also out of scope for our work. We assume a sequential organization of data in the file, which is typical for many applications such as computational astronomy and graph algorithms. We believe that the concepts and techniques of this paper can be used for reading HDF files as well.

\section{Rationale, API and Design}

\ellide{
Our design of the CkIO input library is driven by the following goals and constraints: 
\begin{itemize}
    \item Selective reads prefetch only the necessary data, typically before user requests specific reads. To support this, we require that the user (i.e. each client task) specifies what portion of the file they would like to read from. 
    \item The number of chares that perform file input is chosen independent of the number of application chares (clients) that need the data. This allows users to chose the best decompositions for various file inputs and computations. 
    \item We support concurrent background work, allowing applications to stay active while I/O completes, effectively overlapping the I/O time with useful computations.
    \item  We support the ability for the application to migrate tasks across processors, while holding the file and session handles the opened earlier. 
\end{itemize}
}
We recap and elaborate on the motivations we discussed in the introduction, to provide guideline and rationale for our design. 

Since overdecomposed systems have a large number of persistent tasks (client chares), doing explicit input from each is inefficient. So we need a designated set of I/O agents (chares in a chare array) for actually performing I/O, separate from the application tasks. The size of this chare array (i.e. number of I/O agents) should be selected independent of the number of client chares. 

If the file operations were to begin just when the application needs input data, our ability to overlap useful work with file input will be limited. Instead, our design must allow pre-fetching of file data to the extent possible. 

Task-based systems require that a processor is never blocked on any operation, although individual tasks may block on their data/message dependencies. If CkIO operations or any component activity blocks the whole processor, it will lead to unacceptable performance loss and possible correctness issues due to loss of progress guarantees. So, we must support concurrent background work, allowing applications to stay active while I/O completes, effectively overlapping the I/O time with useful computations.

Object migration is a key feature of Charm++, responsible for many of its advantages.
CkIO must support the ability for the application to migrate tasks across processors, while holding the file and session handles they opened earlier.

\ellide{
Our design of the CkIO input library is driven by the following goals and constraints: 
\begin{itemize}
    \item Selective reads prefetch only the necessary data, typically before user requests specific reads. To support this, we require that the user (i.e. each client task) specifies what portion of the file they would like to read from. 
    \item The number of chares that perform file input is chosen independent of the number of application chares (clients) that need the data. This allows users to chose the best decompositions for various file inputs and computations. 
    \item We support concurrent background work, allowing applications to stay active while I/O completes, effectively overlapping the I/O time with useful computations.
    \item  We support the ability for the application to migrate tasks across processors, while holding the file and session handles the opened earlier. 
\end{itemize}
}

\subsection{Read Session}
We introduce the concept of a \textit{read session} in order to create a separate set of chares as file reading agents, and to allow pre-fetching  of file data. 
The read session is a component of the CkIO API; it must be initialized before any read operations begin. 
This is distinct from opening of a file. Given an open file, a read session is started by all the client objects collectively, providing a heads-up to the system about which portion(s) of the file each client plans to read. A read session call specifies the file handle, start byte, and end byte of a single file that the client will eventually read from via one or more read requests.
Given this initial information, CkIO greedily reads in the input data for the section of the file that is relevant to the application. This greedy reading of data is one feature that allows for the overlapping of the I/O with other application level computation and messaging, which will be described in more detail. In addition, when files cannot fit into memory, the read session concept allows the user to read the file chunk-by-chunk (one chunk per session), where each chunk can fit into memory.

Additionally, because the selective pre-fetching maintains the data in memory, this I/O schema is amenable to other forms of parallelism, such as pipeline parallelism. Briefly, imagine a program in which every worker is required to performs some computation on large file in a block-cyclic fashion. That is, if there are $n$ workers, the $i^{th}$ worker will be assigned the ${j^{th}}$ block if $j \equiv i \mod n$. Also assume that a worker must complete the work of one element before consuming their next element. Because the reads in our scheme are non-blocking, with a split-phase callback interface, and can be scheduled by the runtime, the worker can issue the read request for the next element before executing the work for their current element. 
This can be accomplished by using a series of read-sessions, one for each segment of the file corresponding to $n$ workers' blocks. 
This allows for the read to commence while the worker is doing its computation, effectively overlapping communication and computation. 


\subsection{Independent decomposition of file input tasks and application tasks}

\begin{figure}
\begin{subfigure}{.45\textwidth}
  \centering
  
 \includegraphics[width=.8\linewidth]{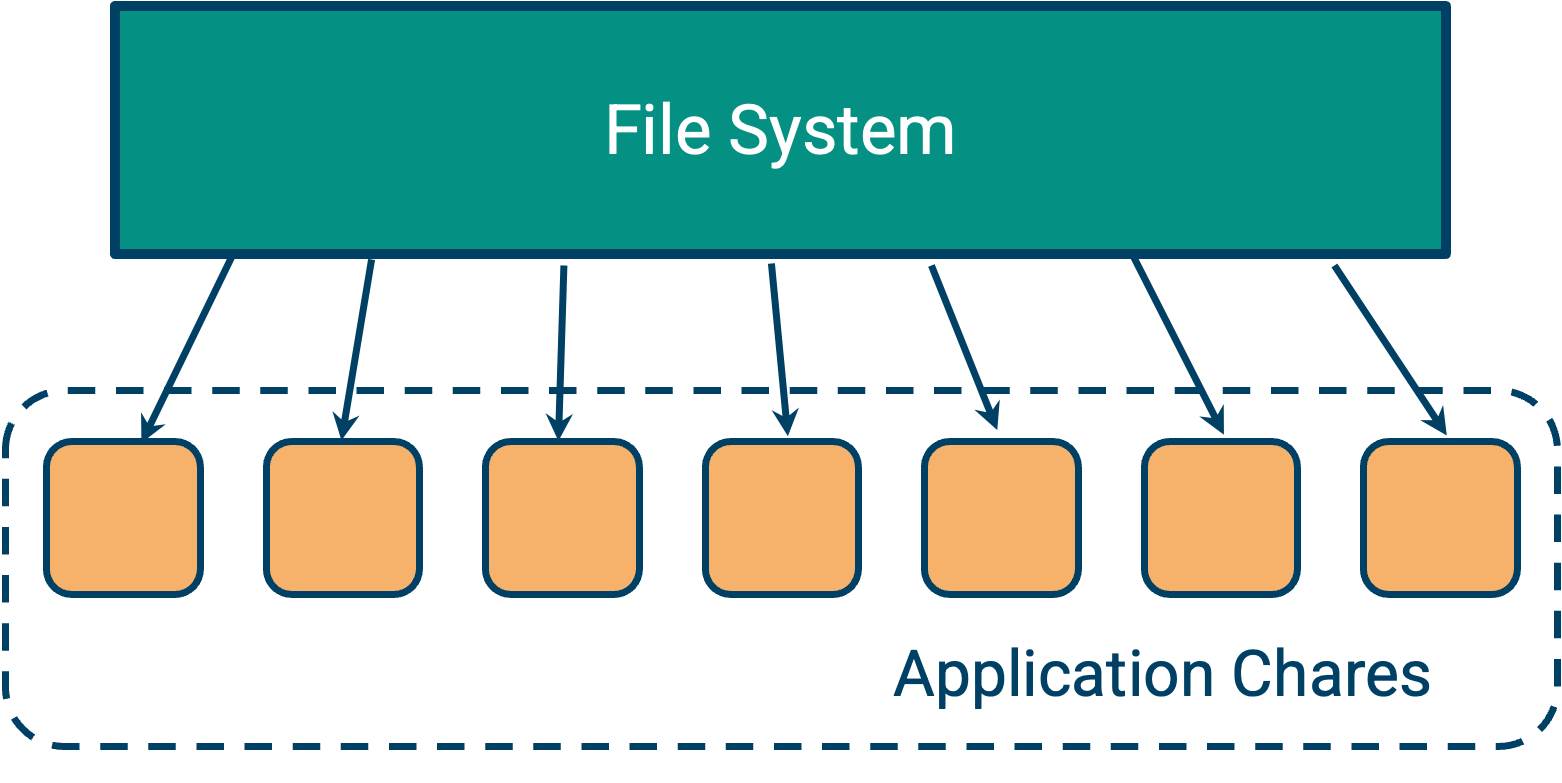}
  \caption{Naive Parallel Input}
  \label{fig:naive}
\end{subfigure}

\vspace{4mm}

\begin{subfigure}{.45\textwidth}
  \centering
  \includegraphics[width=.8\linewidth]{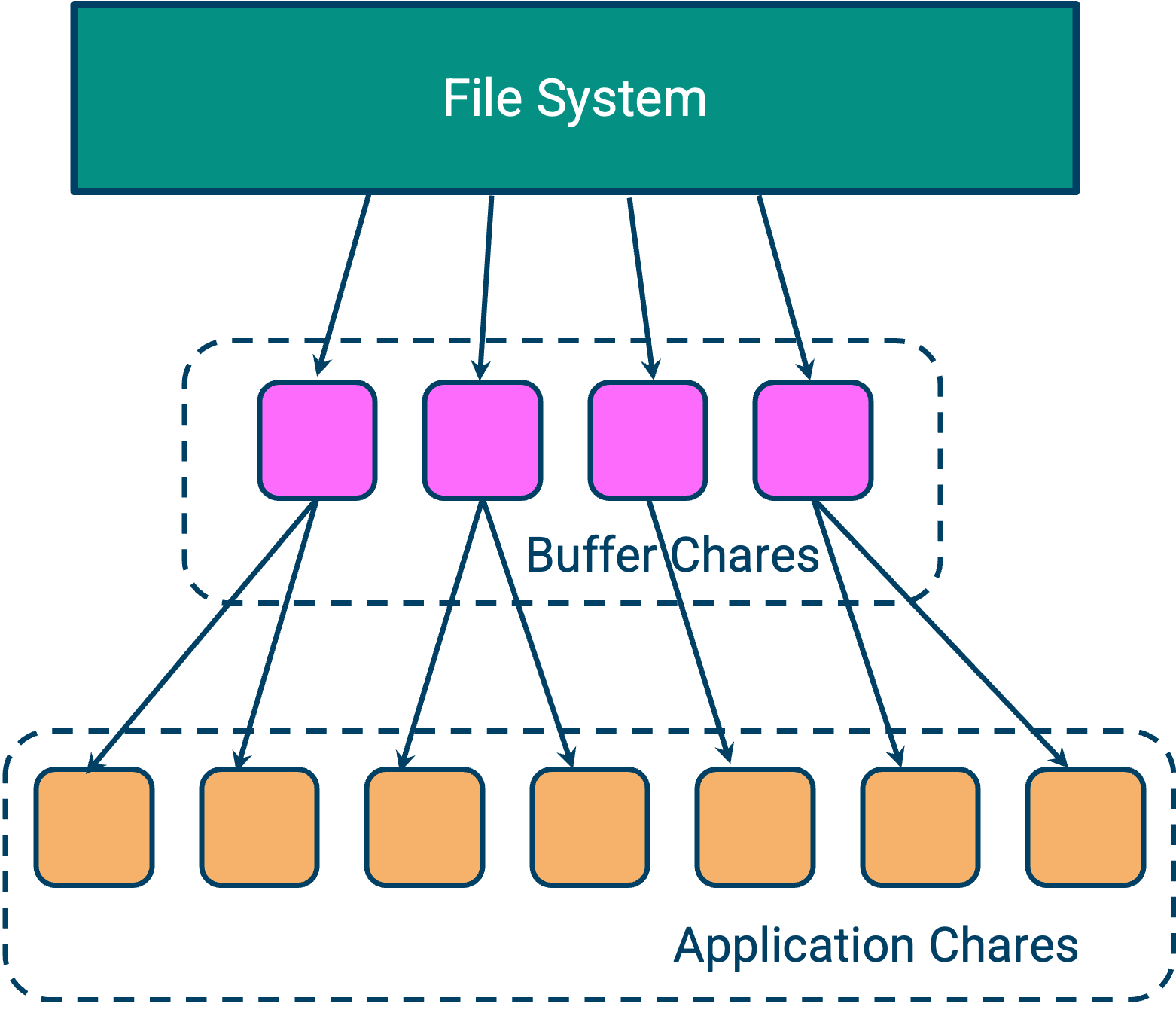}
  \caption{CkIO Input with Buffer Chares}
  \label{fig:bufferChares}
\end{subfigure}
\caption{Schematics of (a) naive parallel input vs (b) input with CkIO. In the naive implementation, application chares interact directly with the file system. With CkIO,  a layer of buffer chares is used to abstract the file system interaction away.}
\label{fig:diagram}
\end{figure}

The basic idea behind the CkIO software architecture, which helps it accomplish its goals of supporting overdecomposition and migratability, is the separation of file input decomposition from application decomposition via a two-phase input implementation. Our library aims to support application developers with performant file input regardless of how they chose to over-decompose their application. To do this, CkIO uses an intermediary chare array responsible for actual file input, called the \textit{buffer chare} array, which sits between the file system and the application chares, or \textit{clients}. Figure~\ref{fig:diagram} shows how the layer of buffer chares abstracts the interaction with the file system from the application.

The buffer chares read disjoint sections of a file that an application will need. The number of buffer chares should be configured by the application (or chosen by the system) to decompose the file-system interaction optimally, depending on a number of factors including the number of PEs, number of nodes, and amount of data to be read. When a client requests data via the CkIO infrastructure, this request is forwarded to and handled by the buffer chare responsible for the relevant data.

\begin{figure}

  \centering
  
 \includegraphics[width=.8\linewidth]{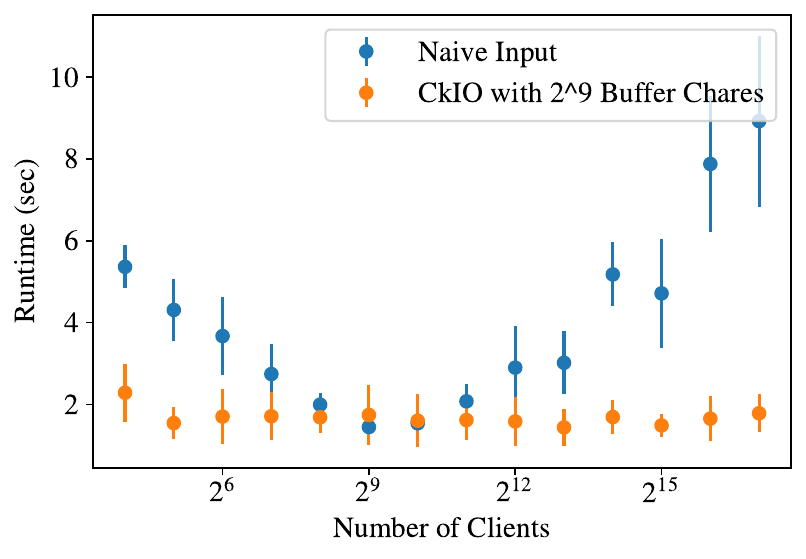}
  \caption{Performance of naive parallel input (where each client directly makes file-system calls) vs input with CkIO reading from a single 4GB file on Bridges2 (16 nodes, 32 tasks per node). As the number of clients vary, CkIO provides consistent performance comparable to the optimal input performance. The vertical bars indicate variability due to file system and compute node contention.}
  \label{fig:naivevsckio}

\end{figure}

By introducing this intermediary layer that sits between the clients and the file system, the actual input performance of reading the data from the file system can be optimized regardless of the client decomposition. This two-phase structure incurs the overhead of additional interconnect data transfer from buffer chare to client, but because the interconnects on most supercomputing clusters are much faster than the I/O disks, the improved input time much outweighs any network cost. Figure~\ref{fig:naivevsckio} shows how CkIO can provide consistent performance regardless of application decomposition, by choosing the number of intermediate chares appropriately.

\subsection{Software Architecture and Implementation}

CkIO's input API is built upon the existing {\em CkIO output} API in order to both have synergy in design between input and output, but also to make it familiar to users of the {\em CkIO output} API if they want to integrate CkIO input into their own program. 

While reading the descriptions  below, please note that a chare is a message-driven object, and a chare group is a collection of chares with exactly one chare per processor. Chare groups are used to implement application-wide capabilities, such as load-balancing, message aggregation, and in the current case, parallel input. A chare group allows client objects on an individual processor interact with its local member, whereas the chares within the group (spread across processors) cooperate to provide the capability. 

    


\begin{figure*}
    \centering
    \includegraphics[width=1\linewidth]{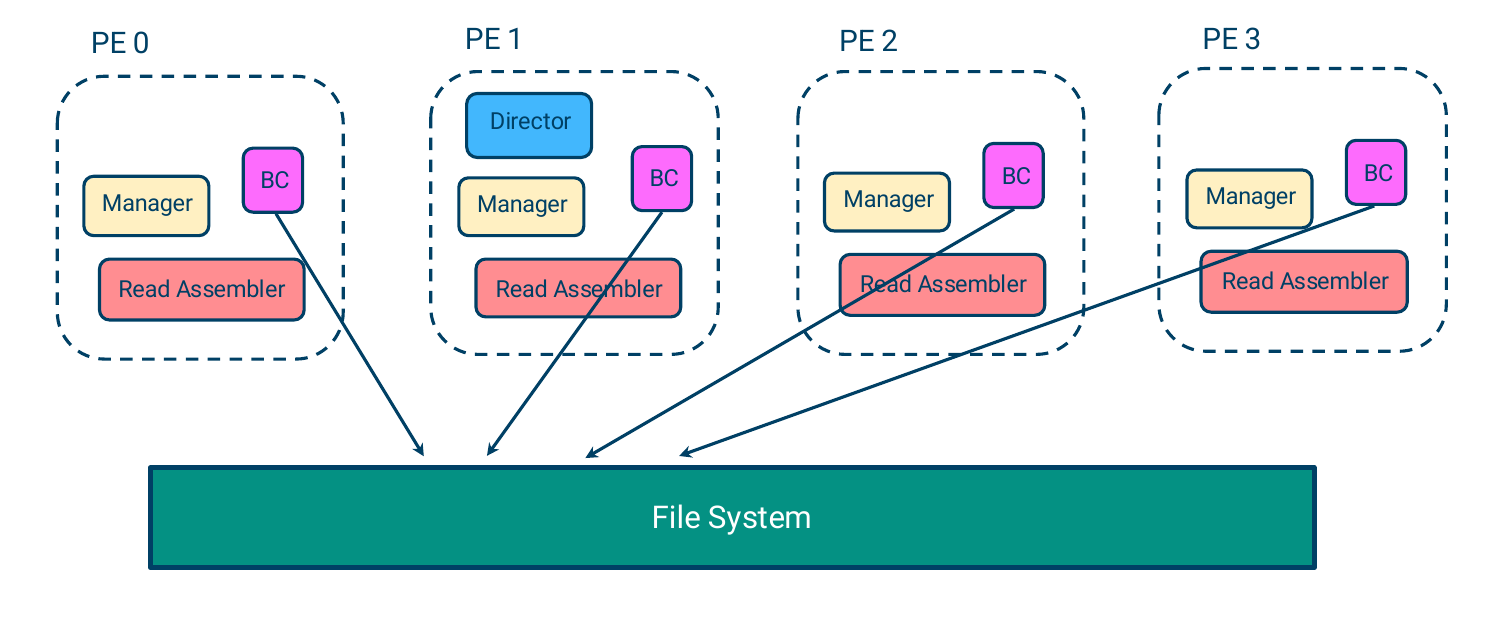}
    
    \caption{Diagram of the CkIO system architecture. Note that the Buffer Chares begin reading on session instantiation, without waiting for client requests. Additionally, the ReadAssemblers are created on instantiation but are not yet active. }
    \label{step_1_system_architecture}
    
\end{figure*}

\begin{figure*}
\begin{subfigure}{1\textwidth}
  \centering
  
 \includegraphics[width=1\linewidth]{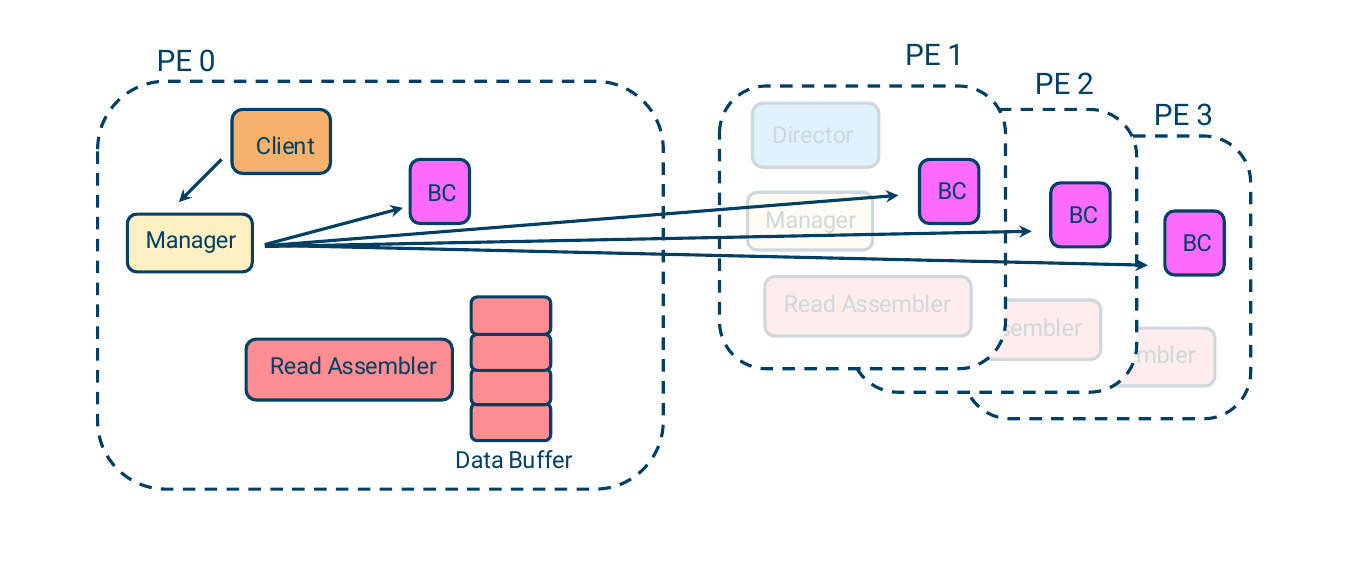}
  \caption{When a client submits a read request to CkIO, this message first passes to the client's local manager. The manager broadcasts the request to all Buffer Chares. }
  \label{fig:send}
\end{subfigure}

\vspace{4mm}

\begin{subfigure}{1\textwidth}
  \centering
  \includegraphics[width=1\linewidth]{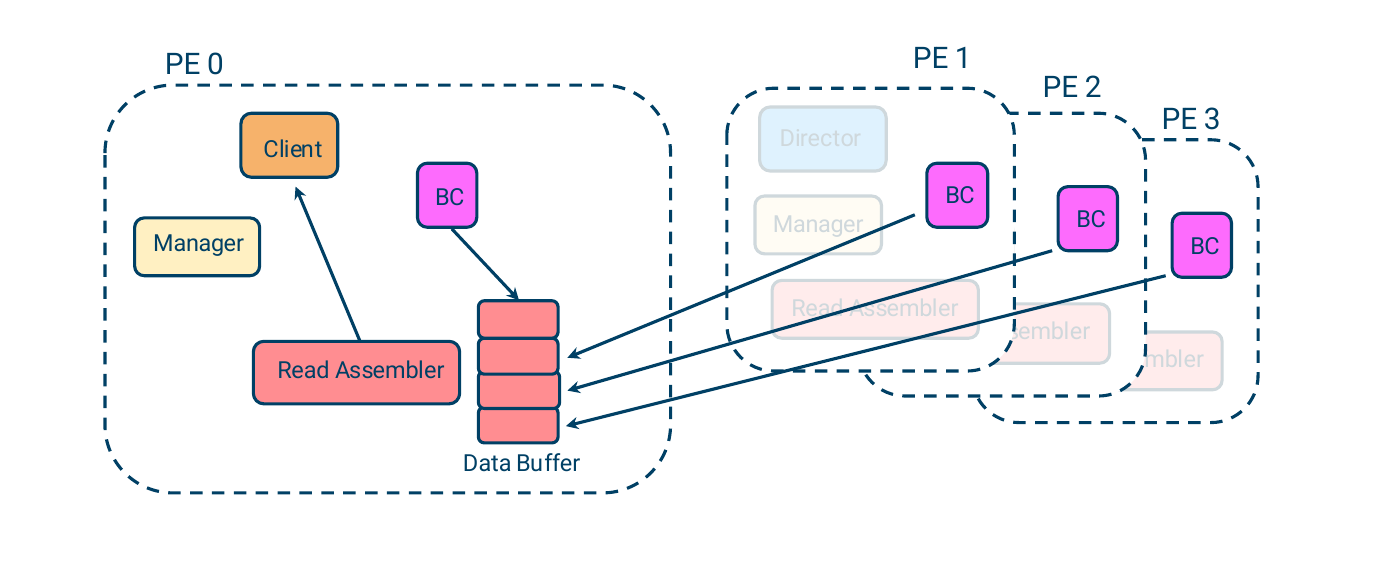}
  
  \caption{After receiving a request, Buffer Chares contribute their relevant data to the Read Assembler on the Client's PE. The Read Assembler assembles received data in a buffer and then passes this to the requesting client.}
  \label{fig:receive}
\end{subfigure}
\caption{Outline of the communication involved when CkIO receives a request from a Client.}
\label{fig:read_architecture}
\end{figure*}

\subsubsection{Director Chare}
The director chare coordinates the actions of the PEs early on in the CkIO input process. When a read session is started, the director broadcasts a message, indicating that a new session has started, to the Manager group, which will be discussed in more detail below.
If global coordination (e.g. sequencing between multiple read sessions of distinct files, to reduce contention in file-system access) is needed, it can be performed by the director chare.

\subsubsection{Manager Group}
The \textit{manager} is a group that is shared with CkIO's output. It is responsible for maintaining a table, mapping each read session to a ReadAssembler group (described in more detail below). Additionally, the manager is responsible for assigning the tags for the zero-copy data transfer used when transferring the data from buffer chares to the clients. This process is described in further detail in the following section. 

\subsubsection{ReadAssembler Group}
The \textit{ReadAssembler} is a group that is responsible for fulfilling the read requests made to the CkIO input library. All the read requests from the clients on a given processor are forwarded to the single ReadAssembler chare object responsible for that PE. Note that each request may require fetching the data from multiple buffer chares in the general case, although given the degree of overdecomposition, it is likely to be limited to just 1 or 2 consecutive buffer chares. The assembler's job to issue the request to the required buffer chares and as the data arrives, to assemble it in the buffer. When a request is fulfilled (i.e. all its pieces have arrived from the buffer chare), it triggers a user-specified callback to continue execution of the read client. This is described in more detail in subsection \ref{API Subsection}. 
\subsubsection{Buffer Chares}
As mentioned in section 3B, buffer chares are the chares that interact directly with the file system and read the file into memory, later to be served over the network. Each buffer chare reads a disjoint section of the file. The number of buffer chares is selected manually by the programmer, based on variables like the file system, file size, and the number of nodes. This is what allows the programmer to optimize the input performance of their application without changing the number of clients. 
Figure \ref{fig:bufferChares} depicts how buffer chares fit into the input model.


To overlap file input with application computation, CkIO performs non-blocking reads. For each buffer chare, we spawn a new OS-level thread whose sole responsibility is to read the section of the read session its buffer chare is responsible for. This will allow the application to continue while the I/O from the buffer chare is being completed. If a read requests occurs before the I/O is completed, that read request will simply be buffered until the I/O is finished; afterwards, the read request will be fulfilled by doing a zero-copy operation to the requesting client's ReadAssembler. We also note that, since the spawned thread is only doing I/O, it will not hinder the performance of the application code. 

The overall system architecture is visualized in Figure \ref{step_1_system_architecture}. The relationships between these different component when a client issues a read request to the system is visualized in Figures \ref{fig:send} and \ref{fig:receive}.

\subsection{Asynchronous-Callback-centric API} \label{API Subsection}
The CkIO input API is designed to allow the Charm++ runtime system to have enough freedom to schedule work optimally. This is accomplished by using asynchronous callbacks, or chare object continuations,  which are a well-supported feature in Charm++. When such as callback is invoked, the system only enqueues the corresponding function or method invocation as a {\em task} on the the specified processor. Combined with the user-space task-scheduler at the heart of Charm++ runtime, this ensures no I/O call blocks an entire processor.  Below, we briefly describe the base CkIO input API that programmers can leverage to get the best input performance in their application.

\begin{itemize}
    \item 
    \texttt{\textit{Ck::IO::\textbf{open}(std::string \textbf{name}, CkCallback \textbf{opened}, Ck::IO::Options \textbf{opts})}} 
    - is used by the user to open a file handle in CkIO. The \textit{opts} variable is a struct that has different fields that can be configured to change how the file is prepared, as well how the user wants subsequent operations to be conducted. For input, the user will be utilizing the \textit{Ck::IO::Options::numReaders}  field, which tells CkIO how many buffer chares to use when opening a read session with the file. When all the managers are prepared to handle the file, the user-specified \textit{opened} callback is invoked, returning a message containing a \textit{Ck::IO::File} handle to the user. 
    \item \texttt{\textit{Ck::IO::\textbf{startReadSession}(Ck::IO::File \textbf{file}, size\_t \textbf{bytes}, size\_t \textbf{offset}, CkCallback \textbf{ready})}} - used to initiate the prefetch of a section of a seekable file into memory by making the buffer chares read \textit{bytes} bytes, starting from the \textit{start\_byte} byte in the file. The file input itself is asynchronous, allowing for it to be overlapped with computation, as described later. Once all the buffer chares have finished initiating their read, the user specified \textit{ready} callback is invoked to return a \textit{Ck::IO::Session} handle to the user. Note that with this asynchronous callback, the system is free to execute other computational work from the time the call is made to the time the session handle is returned. 
    \item \texttt{\textit{Ck::IO::\textbf{read}(Ck::IO::Session \textbf{session}, size\_t \textbf{bytes}, size\_t \textbf{offset}, char* \textbf{data}, CkCallback \textbf{after\_read})}} - This is the function by which application chares will read, giving CkIO an offset and number of bytes with respect to the overall file the \textit{session} corresponds to. The data will be stored in \textit{data}, which is buffer passed in by the user. When the read is complete, CkIO will invoke the user-specified asynchronous callback \textit{after\_read} with the result. This callback is typically a method in the client chare object, which is scheduled as an asynchronous task. I.e.  when the input data is available, a task is only enqueued in the system's task queue for the continuation of application logic that depends on this specific {\em read}. Thus, until the input data is available, the system can continue scheduling other computational tasks. 
   
    \item \textit{\texttt{Ck::IO::\textbf{closeReadSession}(Ck::IO::Session \textbf{read\_session}, CkCallback \textbf{after\_end})}} - This allows the user to clean up the buffer chares and the memory associated with \textit{read\_session} on all the manager chares. Once this is done, the user specified \textit{after\_end} callback is invoked.  
    \item \texttt{\textit{Ck::IO::\textbf{close}(Ck::IO::File \textbf{file}, CkCallback \textbf{closed})}} - Closes the file across all of the PEs. 
\end{itemize}

\section{Evaluation}
\subsection{Micro-benchmark Evaluation}
In this section, we utilize three different microbenchmarks in order to evaluate different aspects of CkIO's design. 
Section \ref{CHaNGaExperiment}
evaluates CkIO on a real world cosmology application. 
All of these experiments were executed on the Bridges2 supercomputer at the Pittsburgh Supercomputing Center \cite{Bridges2}, using the RM nodes. These nodes have 256GB RAM, use the Mellanox ConnectX-6-HDR Infiniband 200Gb/s Adapter as the network, and use Bridges' Lustre PFS, Ocean \cite{Bridges2}. 
\subsubsection{Disjoint reads: MPI vs CkIO}
\begin{figure}

  \centering
  
 \includegraphics[width=.8\linewidth]{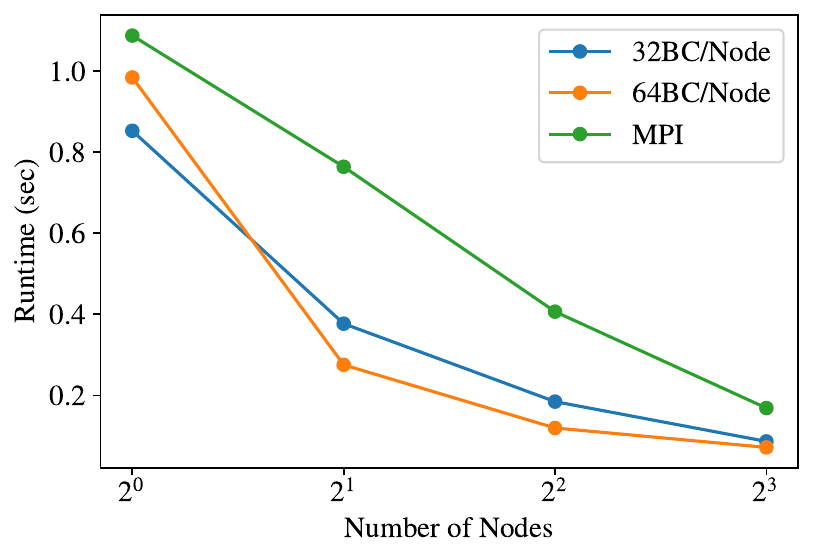}
  \caption{Comparison between MPI-IO and CkIO (with 32 and 64 buffer chares per node) reading a 1GB file with 32 ranks per node on Bridges2.}
  \label{fig:mpi}

\end{figure}

The goal of this experiment is to see how CkIO raw input performance scales compared to other standard HPC I/O frameworks, namely MPI I/O. In this experiment, there is a basic MPI I/O program and a Charm++ program that uses CkIO. In the MPI I/O program, each rank uses collective input to read equal, disjoint chunks of an input file. This is similar in the CkIO program, where there is both $1$ buffer chare and client chare per PE. Figure \ref{fig:mpi} shows a direct comparison between the two I/O implementations. For the MPI I/O benchmark, we used OpenMPI's collective input.

We see that CkIO input does remain better than OpenMPI's collective input from 1 to 8 nodes, with 32 PEs per node, on average. We also note that because of how easy it is to tune the number of buffer chares in CkIO, a user can easily tune their I/O to get the best performance at any scale. 

\subsubsection{Computation Overlap}
In a task-based system such as Charm++, it is desirable that, when tasks performing file input (input tasks) are in progress, other tasks that are not dependent on the input tasks be allowed to proceed concurrently. To evaluate the ability of CkIO to efficiently overlap computation with file input, we use two distinct metrics: (1) the impact of computation/input overlap on total runtime, and (2) how much background work can be completed during the duration of the file input. 
\begin{figure}

  \centering
  
 \includegraphics[width=0.8\linewidth]{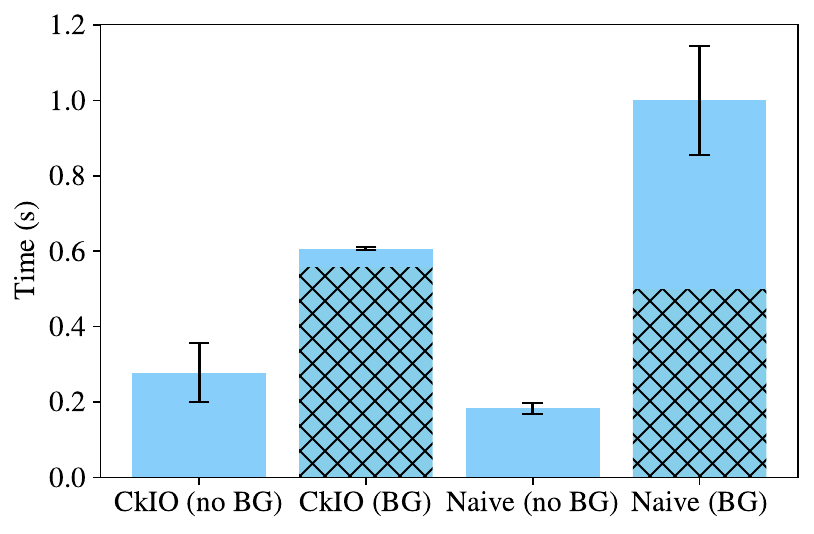}
   \caption{Runtime comparison between CkIO input and naive Charm-based input, with and without fixed background work. Results collected on four nodes of Bridges2, with two cores per node and eight PEs total. Each experiment involves eight clients and eight buffer chares. All approaches read a 1GB file and the data points with background work launch a chare group to perform a fixed amount of background work (hatched) concurrently. Each data point denotes the average recorded runtime over 3 runs. }
   \label{fig:bg-small}

\end{figure}
The first benchmark (Figure \ref{fig:bg-small}) compares the total runtime of the naive file input to file input with CkIO, with and without a fixed amount of background computation. The background work consists of one chare on each PE that iterates over a fixed-duration loop with approximately 10 microseconds of computation per iteration. At the end of every iteration, each chare in the group yields control to the Charm scheduler, giving the runtime a chance to check if the I/O has completed. The execution is completed when both the specified iterations of background computation work are finished and I/O is complete. 

In the naive implementation of file input, chares read directly from the file system and block their respective cores in the process, delaying the background computation. This is visible in Figure \ref{fig:bg-small}, where the naive benchmark runtime more than doubles when background work is added, compared to the baseline with no background work. In the same situation, CkIO is able to overlap input and computation such that a large percentage of the overall runtime is spent in background work, and the runtime only increases slightly when concurrent background work is added.

\begin{figure}

  \centering
  
 \includegraphics[width=0.8\linewidth]{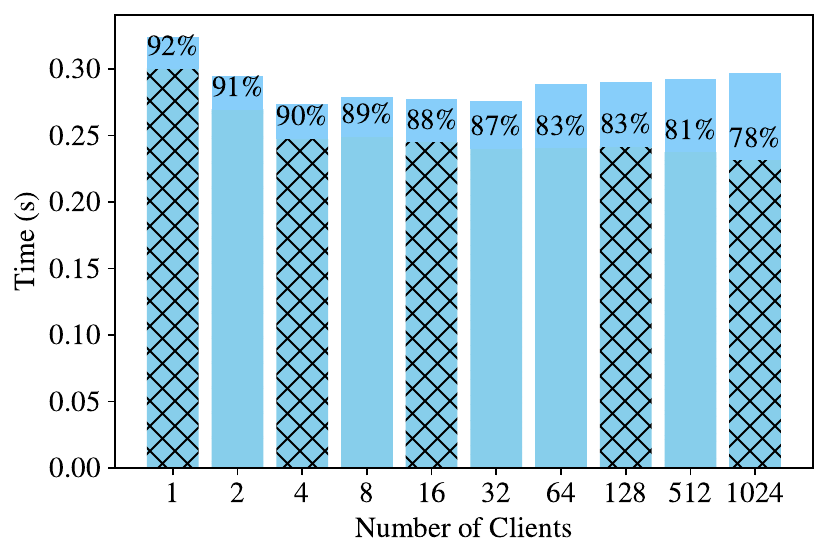}
   \caption{Execution time and percentage of time spent on background work during a read. These results were collected on 4 nodes of Bridges2 with 2 cores per node (8PEs total), using 8 Buffer Chares. 
   }
   \label{fig:bg-variable}

\end{figure}
To further analyze exactly how much overlap is achieved, consider a similar benchmark: 
we create a chare array $A$ and chare group $B$. $A$ is responsible for reading in the entire file, while $B$ is responsible for doing background work until $A$ has finished reading the file. This is then used to measure what percentage of the time $A$ spent reading the file was used by group $B$ doing background work. 
Figure \ref{fig:bg-variable} presents the results. We see that up to1024 clients (64 clients per PE), over 75\% of the input time is utilized for background work.  Beyond that, the extra work of managing I/O for many clients starts impacting the background work fraction. This is due to the fact that each of these clients sends a request to a given buffer chare, and a buffer chare then handles all requests serially. This increased message congestion eventually limits the runtime's ability to schedule background work. Since typical applications use fewer than 64 chares (clients) per PE, this performance is adequate. In future work, it would be interesting to experiment with different threading and buffering strategies to further increase this overlap.

\subsubsection{Migratability}

Charm++ allows its tasks (i.e. the chare objects) to migrate across processors under the control of an adaptive Runtime System (RTS). The RTS may migrate objects around for dynamic load balancing, for energy optimization \cite{oneSystemToRuleThemAll}, for supporting resource elasticity in cloud environments, or for tolerating faults. 
The question arises: if a chare has opened a file, started a read session and then does a series of read operations within that session, but is migrated to a different processor/node between two reads, what happens? We support this scenario by using the virtual proxy for the client, a mechanism Charm++ supports, in the callback for the read (instead of the processor or process rank). With the object location management supported by Charm++, this ensures that new read requests arising after migration are correctly supported. 

\begin{figure}

  \centering
  
 \includegraphics[width=0.8\linewidth]{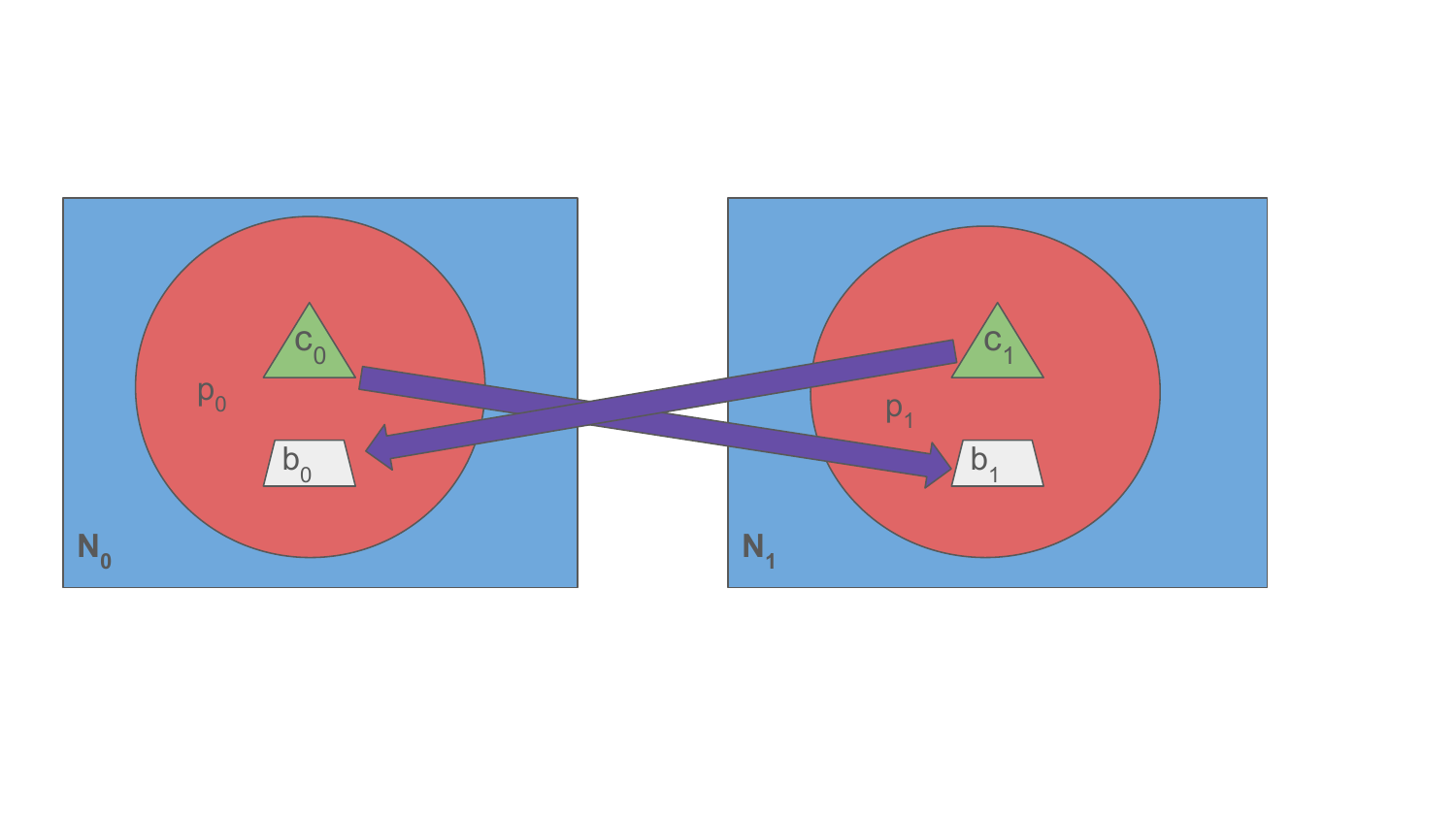}
   \caption{Diagram of the start of the experiment. The arrows indicate that the clients want data belonging to buffer chares on different nodes which will take longer to retrieve.}
   \label{fig:pre_migration_diagram}
 \includegraphics[width=0.8\linewidth]
 {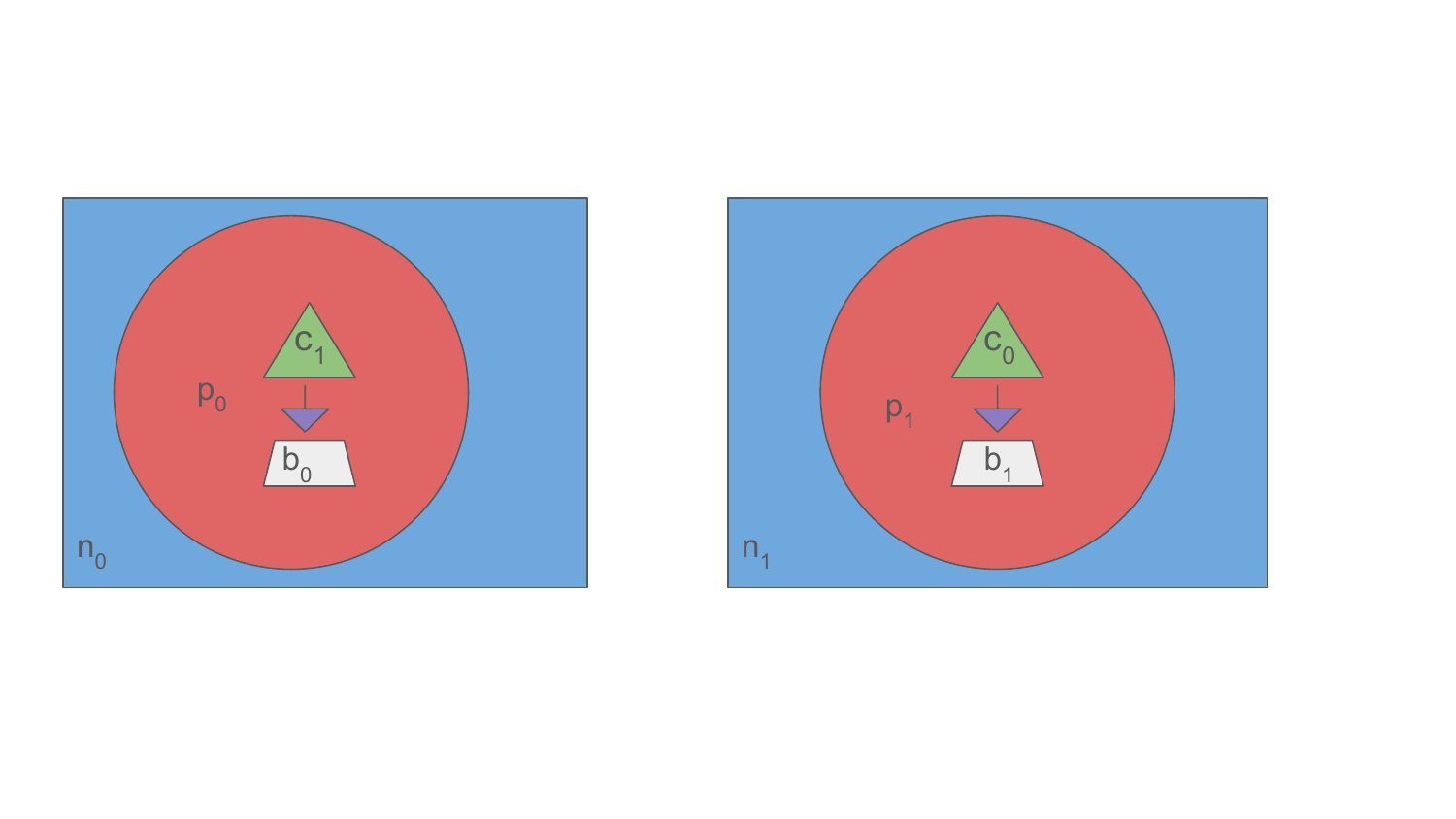}

  \caption{Diagram of the migration experiment after a migration has taken place and the clients want to read data. Now that migration has taken place, both the corresponding client and buffer chares are on the same PEs, which means the request no longer has to cross the node boundary. }
  \label{fig:post_migration_diagram}

\end{figure}

We use an experiment to both demonstrate the support for object migration, as well as a potential locality enhancing optimization carried out by the application: 
On two nodes, we have 2 PEs, where PE $0$, or $p_0$ is on node $0$, or $n_0$, and PE $1$, $p_1$, is on node $1$, $n_1$. The user will use $2$ buffer chares, $b_0$ and $b_1$, respectively, to read in the data from an input file. The application will also have two client chares, $c_0$ and $c_1$, respectively. At the start of the program, $b_0$ and $c_0$  resides on $p_0$, while $b_1$ and $c_1$ resides on $p_1$. During the program,  $c_0$  requests the data that belongs to $b_1$, and $c_1$ requests the data that belongs to $b_0$. Notice that the data wanted by each of the clients happens to live in the opposite node. We then migrate the clients such that each client moves to the other PE and then conduct an identical-size read, this time with the data residing on the same PE as the requesting client. Figure \ref{fig:pre_migration_diagram} illustrates the state of the experiment before migration, and \ref{fig:post_migration_diagram} visualizes the state and actions of the experiment after migration. 

\begin{figure}[t!]

  \centering
  
 \includegraphics[width=1.1\linewidth]{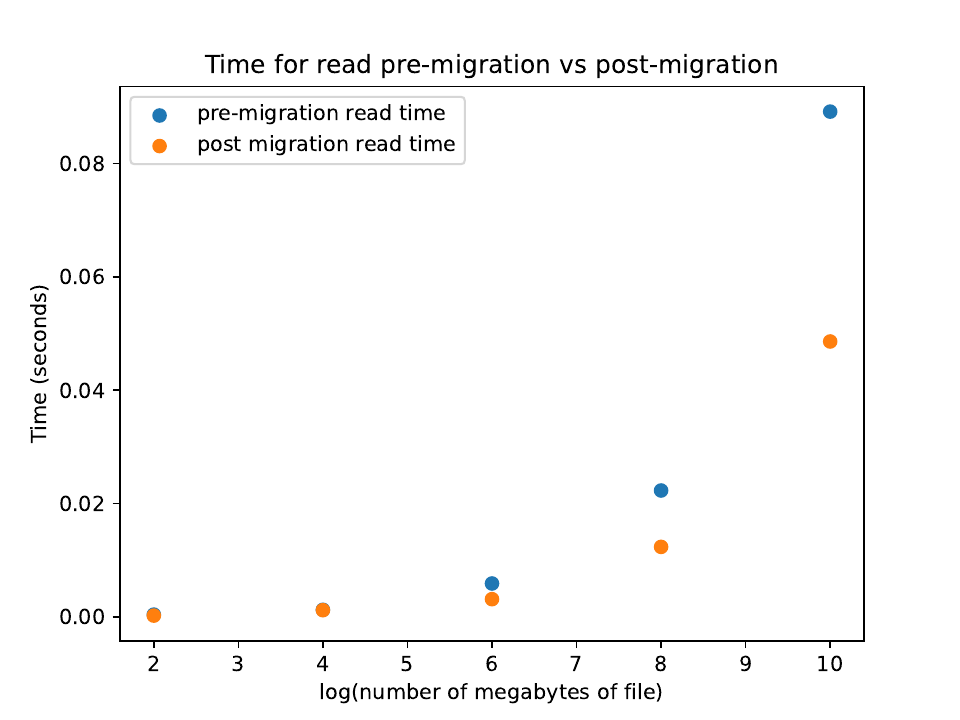}
  \caption{The performance difference in read times before vs after the migration of the client chares. The x-axis corresponds to the log of the number of megabytes of the input file i.e $10$ corresponds to a $1024$MB file. This experiment used two regular-memory nodes, 1 PE per node from Bridges2.}
  \label{fig:migration_times}

\end{figure}

The fact that the program carries out input correctly demonstrates our support for migratability. To study the impact of locality,
Figure \ref{fig:migration_times} illustrates the difference in read speed before migration and after migration as we scale the file size. We plot the largest read time between $c_0$ and $c_1$ as the time taken for $c_0$ and $c_1$ to read in their data. As the file sizes increase, we can see that the disparity between pre-migration read times and post-migration read times increases as well.

This experiment also illustrates the design freedom CkIO provides to the application. Once the read session data has been read, if the amount of data is large and state of the client chare is small,  it may make sense to "send the work to data" as this experiment does. This migration may continue to benefit over subsequent read sessions as well, if the division of data is similar. At the same time, the runtime system is free to migrate chares during the later computational phase(s) if that is beneficial for load balance. Whether migration during Input as done here is beneficial for the application depends on many application-specific factors; so it is important that CkIO provides this option. 


\subsection{Evaluation on ChaNGa} \label{CHaNGaExperiment}

ChaNGa (Charm N-body GrAvity solver) is a Charm++-based cosmology code for collision-less N-body simulations \cite{CHaNGa}. For I/O, ChaNGa uses the Tipsy file format, a format designed specifically for cosmological N-body simulations \cite{tipsy}. 

ChaNGa is implemented using a Barnes-Hut tree composed of a chare array of ``TreePieces''. The simulation space is divided up so that each TreePiece is responsible for a subset of particles. During the initialization phase of the algorithm, TreePieces collectively read disjoint sections from a single input file corresponding to the initial allocation of particles to TreePieces. After completing the input phase, TreePieces are then responsible for driving the Barnes-Hut computation.

 Users can specify the number of TreePieces at runtime and will likely choose to over-decompose the problem, creating many more TreePieces than there are physical cores. While this over-decomposition will benefit computational efficiency, naively performing file input from each TreePiece can cause file system contention and unsatisfactory input performance - precisely the concern CkIO aims to address. 
 
 To mitigate this issue, the ChaNGa developers initially implemented a custom, application-level, collective input scheme. This involves additional code to select a subset of TreePieces (specifically, one per PE) to perform file input and later redistribute particles to all TreePieces. Each of these input-designated TreePieces creates a TipsyReader object to read from the Tipsy input file and reconstruct the simulation particles. The TipsyReader is implemented using the C++ standard std::ifstream to read from the input file via a streaming API.  

We integrate CkIO into CHaNGa, subsuming the manual optimizations written by the ChaNGa developers with CkIO, providing modularity and separation of concerns. In practice, this involved first modifying the TipsyReader implementation to utilize CkIO instead of standard C++ I/O. Secondly, we modified the ChaNGa code-base to replace the custom logic designating select reader TreePieces with CkIO. CkIO supports the logical view that each TreePiece participates in the file input, while making it easy to tune the number of intermediary readers for the workload.

\begin{figure*}
\centering
\begin{subfigure}{.5\textwidth}
  \centering
  \includegraphics[width=.8\linewidth]{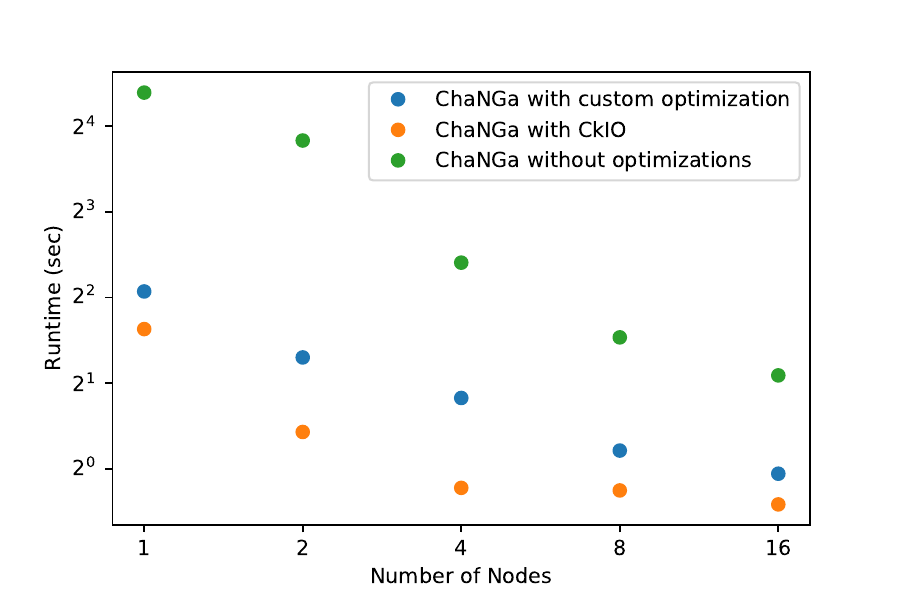}
  \caption{ChaNGa under three input implementations}
  \label{fig:changa-runtimes}
\end{subfigure}%
\begin{subfigure}{.5\textwidth}
  \centering
  \includegraphics[width=.8\linewidth]{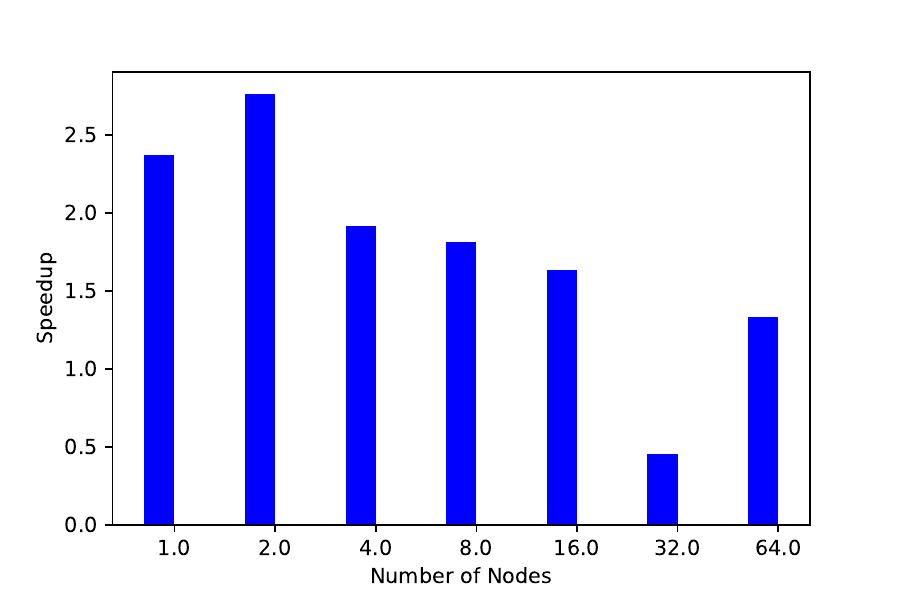}
  \caption{CkIO speedup in ChaNGa}
  \label{fig:changa-speedup}
\end{subfigure}
\caption{Runtime comparisons (a) and speedup (b) of the file input involved in a ChaNGa test code under three IO implementations: 
(1) unoptimized ChaNGa, wherein each TreePiece reads directly from the file system; 
(2) the original ChaNGa implementation, utilizing an application-level optimization that designates one reader TreePiece per PE; and 
(3) ChaNGa with the CkIO implementation of the TipsyReader infrastructure. 
These results were produced on Bridges2 with 32 cores per node, a 1GB input file, and $2^{16}$ ChaNGa TreePieces. The speedup plot visualizes the speedup of ChaNGa with CkIO (3) over the manually optimized ChaNGa implementation (2). Note that while runtimes plotted are the mean, the speedup plotted compares the best iteration (min) of each implementation, to avoid capturing file system irregularities.}
\label{fig:changa}
\end{figure*}

The performance difference of CHaNGa using hand-optimized I/O code compared to using CkIO is illustrated in Figure \ref{fig:changa-runtimes}, in addition to a baseline of ChaNGa without the application-level Tipsy optimizations.
Figure \ref{fig:changa-speedup} shows the speedup of the CkIO implementation over the manually optimized implementation. As an additional datapoint, when running the benchmark on 64 nodes, CkIO shows around a 1.3x speedup, consistent with the trend plotted. These results reaffirm that CkIO provides all the benefits as the hand-optimized ChaNGa implementation, with two additional benefits: the first is apparent in the slight performance improvement provided by CkIO's streaming implementation. Furthermore, the CkIO API provides modularization and abstraction allowing easy tuning of parameters, such as the number of intermediary readers. In comparison to the hand optimized ChaNGa implementation, which was built specifically to have one reader per PE, CkIO enables the user to adapt to environments where one reader per PE may not be optimal, depending on file size, system configuration, and other factors.

Note that while ChaNGa is a good use case for mitigating file system contention due to over-decomposition, it does not provide an opportunity for computation/file input overlap because all input is done prior to any computation.

\section{CkIO Execution Time Analysis}
Lastly, in this section, we seek to understand the major components of CkIO that contribute to the total execution time. Within our system, there are 3 main sources that we wish to study: 
\begin{itemize}
    \item I/O overhead
    \item Data permutation overhead
    \item Overdecomposition overhead
\end{itemize}
In our analysis, we base our study on the disjoint reads program mentioned in Figure \ref{fig:naivevsckio}. 
\subsection{I/O}
In our experiments, we see that the disjoint reads program is I/O bound, with most of the execution time being spent in the application chares waiting for the buffer chares to finish reading their disjoint sections of the file. Using the same setup mentioned in Figure \ref{fig:naivevsckio}, with $2^{9}$ buffer chares and application chares, the extra time spent redistributing the data from buffer chares to clients is minimal. 
\subsection{Data Permutation}
Another major component of our system is the transfer of data from the buffer chare to a client. As seen in Figure \ref{fig:permutation_io}, the time to to transfer data across the network is orders of magnitude faster than reading it from disk. We see this in Figure \ref{fig:naivevsckio}, where the time using $2^{9}$ buffer chares with $2^{9}$ application chares only takes 20 \% more time than the naive charm experiment, with the data permutation amounting to about 0.296 seconds. This illustrates the time that data permutation adds, as the only difference between this experiment and the naive reads is the extra step of sending the necessary data to the application chares.
\subsection{Overdecomposition}
The third component that contributes to total execution overhead is overdecomposition overhead. Overdecomposition overhead occurs when there are a large amount of actors, each requesting data, that results in more time being spent on actions other than permuting the data or I/O. From Figure \ref{fig:naivevsckio}, we see that the time to complete the program remains relatively stable, even up to 256 clients per PE. This suggests that CkIO contains relatively minimal overdecomposition overhead up to large overdecomposition factors. 
\section{Future Work} \label{Future Work Section}
This work has laid the foundations for research into details of HPC input in over-decomposed task-based systems.  We next describe areas that can be further explored to better understand input performance in these systems. 
\subsection{Buffer Chare Selection Policy}
Thus far, through experimentation, the user is able to easily tune the number of buffer chares being used in order to optimize the I/O performance of their application. However, ideally we would like the CkIO library to optimize the number of buffer chares by default based on metrics of the application and machine (such as the file system being used, the interconnect of the machine, number of cores, etc). This would allow for fewer knobs the user is forced to think about and thus the user would only have to focus on optimizing their application code. 
For parallel file output, there has been significant amount work in selecting the number of ``aggregator'' or buffer chares, including the in ROMIO and MPI I/O \cite{numberOfAggregators}
, as well as a search and model based autotuning approach \cite{autotuning}.
We expect this work can be extended to cover our case of overdecomposed clients and emphasizing file input, with its emphasis on dependent tasks. 


\subsection{Topology Awareness and Buffer Chare Placement}
Parallel I/O encounters further challenges in the context of today's wide variety of network topologies and I/O subsystem layouts. Previous work investigates topology and data layout-aware optimizations in the context of I/O abstractions to further improve I/O performance, including the TAPIOCA library, which provides topology-aware two-phase collective I/O for MPI-IO \cite{tapioca}. The intricacies of over-decomposition in a topology-aware I/O abstraction may provide new opportunities for optimization. 
As the migration experiment illustrates, locality of buffer chares plays a significant, although not prohibitive, part in performance. Future work could explore optimal buffer chare placement in light of client chare placement as well as network topology. 

\subsection{Splintered I/O}
Currently, each buffer chare reads the block of data it is responsible for. So, read requests that require only a subsection of the data being read will have to wait until the entire block is read in, which can cause unnecessary increases in the latency of some read requests. For example, if a buffer chare is responsible for a 1GB chunk, but a read request requires only the first 4MB, it will still have to wait for the buffer chare to finish reading the full 1GB before that request can be fulfilled. Instead, if the buffer chare read in 64MB chunks, the chunk in which that read request's data belong to would be available quickly, allowing the request to be fulfilled quicker. Thus, exploring how such splintering I/O might improve the latency of read request access patterns may improve the performance of CkIO and other AMT I/O systems. This issue is complicated if (for example) a buffer chare is serving multiple workers from different processors, and selecting the right sets of ``splinters'', and potential multiple concurrent I/O requests,  will require further exploration. 

\subsection{New application patterns}
Beyond the HPC applications, for which there is years of experience in the community, the new class of applications coming from parallel graph algorithms, data analytics,  agent-based simulations and machine learning present file input and dependency patterns that are yet to be digested and analyzed. As a library like CkIO gets used by task based systems for such applications, we expect new application patterns to emerge, which will require extension, specialization and optimization of the CkIO approach. 

\section{Conclusion} \label{Conclusion Section}

In this paper, we have illustrated the various challenges that over-decomposed AMT systems face when doing file input. We discuss how, by utilizing a 2-layer split-phase I/O scheme based on asynchronous callbacks, we can speed up the input performance of applications considerably, as well as allow for different capabilities, such as speed ups and load balancing that take into account the data that is being read. We implement these ideas in \textit{CkIO Input}, a library that allow for up to 2x faster parallel reads in over-decomposed, task-based systems. \textit{CkIO Input} allows the user to pick a specified number of buffer chares in order to maximize the input performance of an application, while also allowing for efficient overlap of computation and I/O. Additionally, \textit{CkIO Input} supports client object migratability, which is amenable to different strategies to speed up input, as well as allowing for different load-balancing strategies. We  described the implementation of this library in Charm++. Finally, we evaluated CkIO input on three different microbenchmarks, as well as a cosmological N-body simulation. We hope this research spurs further improvements in file input for HPC applications, as well as exploration of various strategies and decompositions schemes for AMT systems performing parallel reads. 

\section{Acknowledgements}
We thank Zane Fink and Ritvik Rao for their feedback on this manuscript. We also thank Bridges 2 administrators and the NSF ACCESS program for use of the Bridges2 machine.

\bibliographystyle{plain}
\bibliography{refs}
\end{document}